\documentclass{aip-cp}

\usepackage[numbers]{natbib}
\usepackage{rotating}
\usepackage{graphicx}
\usepackage{mathrsfs}
\usepackage{amssymb}
\usepackage{graphicx}
\usepackage[normalem]{ulem}
\usepackage[dvips]{color}
\usepackage{bm}
\usepackage{longtable}
\usepackage{times}

\renewcommand\sout{\bgroup \color{red} \ULdepth=-.5ex \ULset}

\def\esym{$E_{\rm{sym}}(\rho)$~}

\renewcommand{\v}[1]{\textbf{#1}}
\renewcommand{\rm}[1]{\textrm{#1}}
\renewcommand{\d}{\mathrm{d}}

\begin{document}

\title{Effects of Neutron-Proton Short-Range Correlation on the Equation of State of Dense Neutron-Rich Nucleonic Matter}
\author[aff1,aff2,aff3]{Bao-Jun Cai}
\author[aff1]{Bao-An Li\corref{cor1}}
\corresp[cor1]{Corresponding author and speaker: Bao-An.Li@tamuc.edu}
\author[aff3]{Lie-Wen Chen}
\affil[aff1]{Department of Physics and Astronomy, Texas A\& M University-Commerce, Commerce,
Texas, 75429, USA}
\affil[aff2]{Department of Physics, Shanghai University, Shanghai 200444, China}
\affil[aff3]{School of Physics and Astronomy and Shanghai Key Laboratory for Particle Physics and Cosmology, Shanghai Jiao Tong University, Shanghai 200240, China}
\maketitle

\begin{abstract}
The strongly isospin-dependent tensor force leads to short-range correlations (SRC) between neutron-proton (deuteron-like) pairs much stronger than those between proton-proton and neutron-neutron pairs. As a result of the short-range correlations, the single-nucleon momentum distribution develops a high-momentum tail above the Fermi surface. Because of the strongly isospin-dependent short-range correlations, in neutron-rich matter a higher fraction of protons will be depleted from its Fermi sea and populate above the Fermi surface compared to neutrons. This isospin- dependent nucleon momentum distribution may have effects on: (1) nucleon spectroscopic factors of rare isotopes, (2) the equation of state especially the density dependence of nuclear symmetry energy, (3) the coexistence of a proton-skin in momentum space and a neutron-skin in coordinate space (i.e., protons move much faster than neutrons near the surface of heavy nuclei). In this talk, we discuss these features and their possible experimental manifestations. As an example, SRC effects on the nuclear symmetry energy are discussed in detail using a 
modified Gogny-Hartree-Fock (GHF) energy density functional (EDF) encapsulating the SRC-induced high momentum tail (HMT) in the single-nucleon momentum distribution. 
\end{abstract}

\section{Single-nucleon momentum distribution function encapsulating SRC effects}\label{sec2} 
It is well known that the SRC leads to a high (low) momentum tail
(depletion) in the single-nucleon momentum distribution function
denoted by $n_{\v{k}}^J$ above (below) the nucleon Fermi surface in
cold nucleonic matter\,\cite{Bethe,Ant88,Arr12,Cio15,Hen16x}.
Significant efforts have been made in recent
years both theoretically and experimentally to constrain the isospin-dependent
parameters characterizing the SRC-modified $n_{\v{k}}^J$ in
neutron-rich nucleonic matter.
In particular, it has been found via analyzing electron-nucleus scattering data that the percentage of
nucleons in the HMT above the Fermi surface is as high as about
28\%$\pm$4\% in symmetric nuclear matter (SNM) but decreases gradually to about only
1\%$\sim$2\% in pure neutron matter (PNM). On
the other hand, the predicted size of the HMT still depends on the
model and interaction used. For instance, the self-consistent
Green's function (SCGF) theory using the AV18 interaction predicts a
11\%$\sim$13\% HMT for SNM at saturation density and a 4\%$\sim$5\%
HMT in PNM\,\cite{Rio09}.

For completeness and the ease of the following discussions, we first
briefly describe the SRC-modified single-nucleon momentum
distribution function encapsulating a HMT constrained by the
available SRC data that we shall use in this work. The
single-nucleon momentum distribution function in cold asymmetric nuclear matter (ANM) has the
following form\,\cite{Cai15,Cai16a,Cai16b,Cai16c}
\begin{eqnarray}\label{MDGen}
n^J_{\v{k}}(\rho,\delta)=\left\{\begin{array}{ll}
\Delta_J,~~&0<|\v{k}|<k_{\rm{F}}^J,\\
&\\
\displaystyle{C}_J\left({k_{\rm{F}}^{J}}/{|\v{k}|}\right)^4,~~&k_{\rm{F}}^J<|\v{k}|<\phi_Jk_{\rm{F}}^J.
\end{array}\right.
\end{eqnarray}
Here, $k_{\rm{F}}^J=k_{\rm{F}}(1+\tau_3^J\delta)^{1/3}$ is the Fermi
momentum where $k_{\rm{F}}=(3\pi^2\rho/2)^{1/3}$, $\tau_3^{\rm{n}}=+1$ and $\tau_3^{\rm{p}}=-1$, respectively,
$\delta=(\rho_{\rm{n}}-\rho_{\rm{p}})/(\rho_{\rm{n}}+\rho_{\rm{p}})$ is the isospin asymmetry,  and $\phi_J$ is a high-momentum cut-off parameter. 
The parameters involved depend on the isospin asymmetry and satisfy the normalization condition \cite{Cai16c}.
The above form of $n^J_{\v{k}}(\rho,\delta)$ was found consistent
with the well-known predictions of microscopic nuclear many-body
theories\,\cite{Bethe,Ant88,Arr12,Cio15} and the recent experimental
findings\,\cite{Hen16x}. This form of $n^J_{\v{k}}(\rho,\delta)$ has been applied to address several issues
regarding the HMT effects recently in both nuclear physics and astrophysics.

The parameters $\Delta_J$, $C_J$ and $\phi_J$ are assumed to depend
linearly on $\delta$ based on predictions of microscopic many-body
theories\, $Y_J=Y_0(1+Y_1\tau_3^J\delta)$\,\cite{Cai15}. The amplitude ${C}_J$
and high-momentum cutoff coefficient $\phi_J$ determine the fraction
of nucleons in the HMT via
$
x_J^{\rm{HMT}}=3C_{{J}}\left(1-\phi_J^{-1}\right).
$
Moreover, the normalization condition between the density $\rho_J$
and the distribution $n_{\v{k}}^J$, i.e.,  $
[{2}/{(2\pi)^3}]\int_0^{\infty}n^J_{\v{k}}(\rho,\delta)\d\v{k}=\rho_J={(k_{\rm{F}}^{J})^3}/{3\pi^2}
$ requires that only two of the three parameters, i.e., ${C}_J$,
$\phi_J$ and $\Delta_J$, are independent. Here we choose the first
two as independent and determine the $\Delta_J$ by
$
\Delta_J=1-3{C}_J(1-\phi_J^{-1})=1-x_J^{\rm{HMT}}.
$
Meanwhile, the adopted ${C}/{|\v{k}|^4}$ shape of the HMT both for
SNM and PNM is strongly supported by recent studies both
theoretically and experimentally. It is
interesting to point out that the $|\v{k}|^{-4}$ form of the HMT is
also found in Bose system theoretically and experimentally, indicating a very general
feature of the HMT. For comparisons, we use two HMT parameter sets.
The $n_{\v{k}}^J$ adopting a 28\% HMT in SNM and a 1.5\% HMT in
PNM is abbreviated as the HMT-exp set, and that
adopting a 12\% HMT in SNM and a 4\% HMT in PNM\,\cite{Rio09} as the
HMT-SCGF set\,\cite{Cai16c}. Moreover, the model using a step
function for the $n_{\v{k}}^J$ is denoted as the free Fermi gas
(FFG) set as a reference. As discussed in more details in
ref.\,\cite{Cai16c}, the HMT parameters in the HMT-exp (HMT-SCGF)
parameter set are $\phi_0\approx2.38$ ($\phi_0\approx1.49$),
$\phi_1\approx-0.56$ ($\phi_1\approx-0.25$), $C_0\approx0.161$
($C_0\approx0.121$), and
$C_1\approx-0.25$ $(C_1\approx-0.01)$, respectively.\\

\section{Incorporating SRC effects in Gogny Hartree-Fock energy density functionals} 
In most studies of heavy-ion collisions using transport models, one
parameterizes the energy density functionals (EDFs) and determine their parameters by reproducing
empirical properties of SNM at the saturation density $\rho_0$, a
selected value of symmetry energy $E_{\rm{sym}}(\rho_0)$ and its
slope $L\equiv[3\rho\d E_{\rm{sym}}(\rho)/\d\rho]_{\rho_0}$ as well
as main features of nucleon optical potentials extracted from
analyzing nucleon-nucleus scatterings, such as the isosclar and
isovector nucleon effective masses and their asymptotic values at
high momenta at $\rho_0$, etc., see, e.g., ref.\,\cite{PPNP} for
detailed discussions.  For example, using a modified Gogny-type
momentum-dependent interaction (MDI) \cite{Gogny,Gal87,Das03,Che05,Che14}, a modified GHF-EDF in terms of
the average energy per nucleon $E(\rho,\delta)$ in ANM at density
$\rho$ and isospin asymmetry $\delta$ can be written as
\begin{eqnarray}\label{EDF}
&&E(\rho,\delta)=\sum_{J=\rm{n,p}}\frac{1}{\rho_J}\int_0^{\infty}\frac{\v{k}^2}{2M}n_{\v{k}}^J(\rho,\delta)\d\v{k}
+\frac{A_\ell(\rho_{\rm{p}}^2+\rho_{\rm{n}}^2)}{2\rho\rho_0}
+\frac{A_{\rm{u}}\rho_p\rho_n}{\rho\rho_0}
+\frac{B}{\sigma+1}\left(\frac{\rho}{\rho_0}\right)^{\sigma}(1-x\delta^2)\nonumber\\
&&+\sum_{J,J'}\frac{C_{J,J'}}{\rho\rho_0}\int\d\v{k}\d\v{k}'f_J(\v{r},\v{k})f_{J'}(\v{r},\v{k}')\Omega(\v{k},\v{k}').
\end{eqnarray}
The first term is the kinetic energy while the second to fourth terms are the usual zero-range 2-body and effective 3-body
contributions characterized by their strength parameters $A_\ell,
A_{\rm{u}}$ and $B$ as well as the density dependence $\sigma$ of
the 3-body force\,\cite{Das03,Che05}
\begin{equation} A_\ell=A_\ell^0+\frac{2xB}{1+\sigma},
~~A_{\rm{u}}=A_{\rm{u}}^0-\frac{2xB}{1+\sigma}
\end{equation}
where $x$ controls the competition between the isosinglet and
isotriplet 2-body interactions, and it affects only the slope $L$
but not the $E_{\rm{sym}}(\rho_0)$ by design\,\cite{Das03}. The last
term in Eq. (\ref{EDF}) is the contribution to the Equation of State (EOS) from the
finite-range 2-body interactions characterized by the strength
parameter $C_{J,J}\equiv C_\ell$ for like and
$C_{J,\overline{J}}\equiv C_{\rm{u}}$ for unlike nucleon paris,
respectively, using the notations $\overline{\rm{n}}=\rm{p}$ and
$\overline{\rm{p}}=\rm{n}$. The $f_J(\v{r},\v{k})$ and
$n_{\v{k}}^J(\rho,\delta)$ are  the nucleon phase space distribution
function and momentum distribution function, respectively. In
equilibrated nuclear matter at zero temperature, they are related by
\begin{equation}
f_J(\v{r},\v{k})=\frac{2}{h^3}n_{\v{k}}^J(\rho,\delta)=\frac{1}{4\pi^3}n_{\v{k}}^J(\rho,\delta),~~\hbar=1.
\end{equation}
For example, in the FFG,  $n_{\v{k}}^J=\Theta(k_{\rm{F}}^J-|\v{k}|)$ with $\Theta$ the
standard step function, then $f_J(\v{r},\v{k})=(1/4\pi^3)\Theta(k_{\rm{F}}^J-|\v{k}|)$.

The regulating function $\Omega(\v{k},\v{k}')$\,\cite{Gal87,Das03}
originating from the meson exchange theory of nuclear force normally
has the form of
\begin{equation}\label{Ome}
\Omega(\v{k},\v{k}')=\left[1+\left(\frac{\v{k}-\v{k}'}{\Lambda}\right)^2\right]^{-1}
\end{equation}
where $\v{k}$ and $\v{k}'$ are the momenta of two interacting
nucleons and $\Lambda$ is a parameter regulating the momentum
dependence of the single-particle potential. For applications to
SNM, it is usually determined by fixing the nucleon isoscalar
effective mass at the Fermi surface to an empirical
value\,\cite{Gal87,Das03}. In applying the above formalisms to transport model simulations of
nuclear reactions, the $f_J(\v{r},\v{k})$ and
$n_{\v{k}}^J(\rho,\delta)$ are calculated self-consistently from
solving dynamically the coupled Boltzmann-Uehling-Uhlenbeck (BUU)
transport or molecular dynamics equations for
quasi-nucleons\,\cite{Bert88,Aich91,LCK08}. While in studying thermal
properties of hot nuclei or stellar matter in thermal equilibrium,
the Fermi-Dirac distributions at finite temperatures are used.

Traditionally, one writes the EDF as a sum of kinetic EOS of FFG
plus several potential terms. Before making any applications, the
model parameters of the EDFs are normally fixed by using step
functions for the $f_J(\v{r},\v{k})$ and $n_{\v{k}}^J(\rho,\delta)$
as in a FFG at zero temperature in reproducing properties of nuclei
or nuclear matter in their ground states. In reality, however, since
all nucleons interact with each other in nuclear medium, they
naturally become quasi-nucleons. The normal practice of optimizing
the EDFs puts all effects of interactions into the potential part of
the EDF thus ignores interaction effects on the kinetic energy of
quasi-nucleons. The momentum distribution of these quasi-nucleons in
the ground state of the system considered is not simply a step
function if SRC effects are considered as we discussed in the
previous section.  Here, we separate the total EDF into a kinetic
energy and several potential parts of quasi-nucleons. The
$f_J(\v{r},\v{k})$ and $n_{\v{k}}^J(\rho,\delta)$ with HMTs
constrained by the SRC experiments are used in evaluating both the
kinetic and the momentum-dependent potential parts of the EDF in ANM
at zero temperature. At least for simulating heavy-ion collisions
using transport models, how the total EDFs are separated into their
kinetic and potential parts are important and have practical
consequences in predicting experimental
observables. Interestingly, how the SRC may
affect the symmetry energy, heavy-ion reactions and properties of
neutron stars are among the central issues in our pursuit of
understanding the nature of neutron-rich nucleonic matter. Previous
attempts to incorporate the experimentally constrained
$n_{\v{k}}^J(\rho,\delta)$ and $f_J(\v{r},\v{k})$ with HMT in the
non-relativistic EDF and examine their effects on heavy-ion
collisions and neutron stars were found very
difficult. This is mainly because of the nontrivial
momentum dependence of the $U_J(\rho,\delta,|\v{k}|)$ and the EDF
when the SRC-modified $n_{\v{k}}^J(\rho,\delta)$ and
$f_J(\v{r},\v{k})$ are used. Since one needs to solve 8-coupled
equations simultaneously to obtain self-consistently all model
parameters from inverting empirical properties of ANM and nucleon
optical potentials at $\rho_0$, numerical problems associated with
the momentum integrals in
Eq. (\ref{EDF}) using the original $\Omega(\v{k},\v{k}')$ are very difficult to solve. \\

\section{A surrogate high-momentum regulating function for the MDI energy density functional}\label{sec4}
To overcome the numerical problem mentioned above, a surrogate high-momentum regulating function $\Omega(\v{k},\v{k}')$ that approximates very well the original one while enables all integrals in the EDF and $U_J(\rho,\delta,|\v{k}|)$  to be analytically expressed was proposed recently in ref. \cite{CaiLi}. Perturbatively, if $\Lambda$
is large compared to the momenta scale in the problems under investigation, the $\Omega(\v{k},\v{k}')$ in Eq. (\ref{Ome}) can be expanded as $\Omega(\v{k},\v{k}')\approx 1-{\v{k}^2}/{\Lambda^2}-{\v{k}'^2}/{\Lambda^2}+{2\v{k}\cdot\v{k}'}/{\Lambda^2}$. Using this as a hint, we parameterize the $\Omega(\v{k},\v{k}')$ as
\begin{equation}\label{Omega}
\Omega(\v{k},\v{k}')=1+{a}\left[\left(\frac{\v{k}\cdot\v{k}'}{\Lambda^2}\right)^2\right]^{1/4}
+{b}\left[\left(\frac{\v{k}\cdot\v{k}'}{\Lambda^2}\right)^2\right]^{1/6},
\end{equation}
where $a$ and $b$ are two new parameters. It is interesting to note that this $\Omega(\v{k},\v{k}')$ is invariant under the transformation
$a\rightarrow a/\xi^{3/2}$, $b\rightarrow \xi b$ and $\Lambda\rightarrow \Lambda/\xi^{3/2}$,
indicating that we have the freedom to first fix one of them without affecting the physical results. Here we set $b=2$ and then determine the $a$ and $\Lambda$ using known constraints as we shall
discuss in the following.

The advantages of using this new regulating function is twofold: firstly, the
basically 1/2 and 1/3 power of $\frac{\v{k}\cdot\v{k}'}{\Lambda^2}$ in the second and third term in (\ref{Omega}) is relevant for describing properly the energy
dependence of nucleon optical potential\,\cite{Ham90}; secondly, it enables analytical expressions for the EOS and
$U_J(\rho,\delta,|\v{k}|)$ in ANM.  We notice that the $\Omega$ function is only perturbatively effective at
momenta smaller than the momentum scale $\Lambda$, indicating that the EDF constructed can only be used to a restricted range of momentum/density.
It turns out that the cut-off of the HMT in ANM up to about $3\rho_0$ is significantly smaller than the $\Lambda$ parameter we use here.
The above non-relativistic GHF-EDF is denoted as abMDI in the following.


\begin{table}[tbh]
\caption{Coupling constants used in the three sets (right side) and
some empirical properties of asymmetric nucleonic matter used to fix
them (left side). $b=2$ and $\Lambda=1.6\,\rm{GeV/c}$ are used in this
work. $K_0\equiv K_0(\rho_0),M_0^{\ast}\equiv
M_0^{\ast}(\rho_0),L\equiv L(\rho_0)$.}\label{tab_para}
{\normalsize\begin{tabular}{lr||lrrr}
\hline\hline Quantity& Value & Coupling & FFG~~~~&HMT-SCGF&HMT-exp\\
\hline $\rho_0$ (fm$^{-3}$) & $0.16$ & $A_\ell^0$ $(\rm{MeV})$ &
$-578.7397$&$614.1020$&$307.4366$\\
\hline $E_0(\rho_0)$
$(\rm{MeV})$&$-16.0$&$A_{\rm{u}}^0$ $(\rm{MeV})$&$225.6127$&$711.5675$&$1055.4219$\\
\hline $M_0^{\ast}/M$ &$0.58$&$B$ $(\rm{MeV})$&$517.5297$&$-256.9850$&$-64.5669$\\
\hline $K_0$ ($\rm{MeV}$) &$230.0$&$C_\ell$ $(\rm{MeV})$&$-155.6406$&$-154.2604$&$-129.5643$\\
\hline $U_0(\rho_0,0)$ ($\rm{MeV}$)
&$-100.0$&$C_{\rm{u}}$ $(\rm{MeV})$&$-285.3256$&$-351.5893$&$-587.2980$\\
\hline $E_{\rm{sym}}(\rho_0)$ ($\rm{MeV}$)&$31.6$&$\sigma$&$1.0353$&$0.9273$&$0.6694$\\
\hline $L$ ($\rm{MeV}$)&$58.9$&$a$&$-5.4511$&$-5.0144$&$-4.1835$\\
\hline $U_{\rm{sym}}(\rho_0,1\,\rm{GeV})$ ($\rm{MeV}$)&$-20.0$&$x$&$0.6144$&$0.3703$&$0.1123$\\
\hline\hline
\end{tabular}}
\end{table}


We fix all parameters in the model EDF using empirical properties of
SNM, ANM and main features of nucleon optical potentials at
$\rho_0$. More specifically, for SNM we adopt
$E_0(\rho_0)=-16\,\rm{MeV}$ at the saturation density
$\rho_0=0.16\,\rm{fm}^{-3}$ with $E_0(\rho)=E(\rho,0)$ the EOS of
SNM, its incompressibility $K_0\equiv
[9\rho^2\d^2E_0(\rho)/\d\rho^2]_{\rho_0}=230\,\rm{MeV}$\,\cite{You99,Shl06,Pie10,Che12,Col14},
the isoscalar nucleon k-mass, i.e.,
$M_0^{\ast}(\rho)/M=[1+(M/|\v{k}|)\d
U_0/\d|\v{k}|]^{-1}_{|\v{k}|=k_{\rm{F}}}$, is
selected as $M_0^{\ast}(\rho_0)/M=0.58$, and
$U_0(\rho_0,0)=-100\,\rm{MeV}$. For the isospin-dependent part in
ANM, we adopt $E_{\rm{sym}}(\rho_0)=31.6\,\rm{MeV}$ for the symmetry
energy, $L\equiv L(\rho_0)=58.9\,\rm{MeV}$\,\cite{LiBA13} for the
slope of the symmetry energy and
$U_{\rm{sym}}(\rho_0,1\,\rm{GeV})=-20\,\rm{MeV}$\,\cite{XuJ15} for
the symmetry potential, respectively. Moreover, the value of
$\Lambda$ is constrained to fall within a reasonable range to
guarantee the effect of the high order terms in $\delta$ in the EOS
of ANM mainly characterized by the fourth order symmetry energy,
i.e.,
$E_{\rm{sym},4}(\rho)\equiv24^{-1}\partial^4E(\rho,\delta)/\partial\delta^4|_{\delta=0}$,
is smaller than 3\,MeV at $\rho_0$, to be consistent with
predictions of microscopic many-body theories. Consequently,
$1.40\,\rm{GeV}\lesssim\Lambda\lesssim1.64\,\rm{GeV}$ is obtained
and the study based on  $\Lambda=1.6\,\rm{GeV}$ is used as the
default one. It is worth noting that the single-nucleon potential in
SNM thus constructed is consistent with the global relativistic
nucleon optical potential extracted from analyzing nucleon-nucleus
scattering data\,\cite{Ham90}. Thus, totally five isoscalar
parameters, i.e., $A_{\rm{t}}\equiv
A_\ell+A_{\rm{u}},B,C_{\rm{t}}\equiv C_\ell+C_{\rm{u}},\sigma$ and
$a$ for SNM,  and three isovector parameters, i.e.,
$A_{\rm{d}}\equiv A_\ell-A_{\rm{u}},C_{\rm{d}}\equiv
C_\ell-C_{\rm{u}}$ and $x$ are
all fixed. Details values of these parameters for the three cases using the same set of input physical properties are shown in Tab. \ref{tab_para} .\\

\begin{figure}[h!]
\centering
  \includegraphics[width=9.5cm]{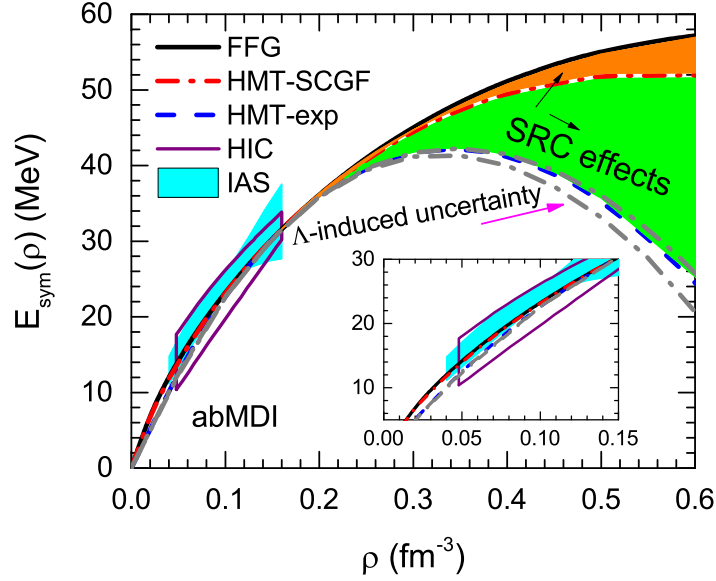}
  \caption{Density dependence of nuclear symmetry energy \esym using the FFG,  HMT-SCGF and HMT-exp parameter set, respectively.
  Constraints on the symmetry energy from analyzing heavy-ion collisions (HIC)\,\cite{Tsa12} and isobaric analog states (IAS)\,\cite{Dan14}
  are also shown for comparisons. The uncertainty range due to the $\Lambda$ parameter is indicated with the gray dash-dot lines for the HMT-exp set.
  Taken from refs. \cite{PPNP,CaiLi}. }
  \label{fig_ab_Esym}
\end{figure}

\section{SRC effects on the density dependence of nuclear symmetry energy}\label{sec4} 
Now we turn to effects of the SRC on nuclear symmetry energy. Shown in Fig.
\ref{fig_ab_Esym} are the results obtained using the FFG, HMT-SCGF
and HMT-exp parameter sets. By construction, they all have the same
$E_{\rm{sym}}(\rho_0)$ and $L$ at $\rho_0$. Also shown are the
constraints on the $E_{\rm{sym}}(\rho)$ around $\rho_0$ from
analyzing intermediate energy heavy-ion collisions
(HIC)\,\cite{Tsa12} and the isobaric analog states
(IAS)\,\cite{Dan14}. Although the predicted \esym using the three
parameter sets can all pass through these constraints, they behave
very differently especially at supra-saturation densities. The
uncertainty of the \esym due to that of the $\Lambda$ parameter is
also shown in Fig. \ref{fig_ab_Esym} for the HMT-exp set with the
gray dash-dot lines. It is seen that the uncertainty is much smaller
than the SRC effect. For example, the variation of the symmetry
energy at $3\rho_0$ owing to the uncertainty of $\Lambda$ is about
2.3\,MeV while the SRC effect is about 14.5\,MeV.  Since the
$\Lambda$ parameter mainly affects the high density/momentum
behavior of the EOS, its effects become smaller at lower densities.
The reduction of the \esym at both sub-saturation and
supra-saturation densities leads to a reduction of the curvature
coefficient $K_{\rm{sym}}\equiv
9\rho_0^2\d^2E_{\rm{sym}}(\rho)/\d\rho^2|_{\rho=\rho_0}$ of the
symmetry energy. More quantitatively, we find that the
$K_{\rm{sym}}$ changes from $-109$\,MeV in the FFG set to about
$-121\,\rm{MeV}$ and $-188$\,MeV in the HMT-SCGF and HMT-exp set,
respectively. It is interesting to stress that this SRC reduction of
$K_{\rm{sym}}$ help reproduce the experimentally measured
isospin-dependence of incompressibility $K(\delta)=
K_0+K_{\tau}\delta^2+\mathcal{O}(\delta^4)$ in ANM where
$K_{\tau}=K_{\rm{sym}}-6L-J_0L/K_0$. The skewness of SNM
$J_0\equiv27\rho_0^3{\d^3E_0(\rho)}/{\d\rho^3}|_{\rho=\rho_0}$ is
approximately $-381$, $-376$ and $-329\,\rm{MeV}$ in the FFG,
HMT-SCGF and HMT-exp set, respectively. The resulting $K_{\tau}$ is
found to change from $-365\,\rm{MeV}$ in the FFG set to about
$-378\,\rm{MeV}$ and $-457\,\rm{MeV}$ in the HMT-SCGF and HMT-exp
set, respectively. The latter is in good agreement with the best
estimate of $K_{\tau}\approx-550\pm 100$\,MeV from analyzing several
different kinds of experimental data currently
available\,\cite{Col14}.

It is also interesting to notice that the SRC-induced reduction of
\esym within the non-relativistic EDF approach here is qualitatively
consistent with the earlier finding within the nonlinear
Relativistic Mean-Field (RMF) theory\,\cite{Cai16b}. Nevertheless,
since there is no explicit momentum dependence in the RMF EDF, the
corresponding reduction of \esym is smaller. Obviously, the
momentum-dependent interaction makes the softening of the symmetry
energy at supra-saturation densities more evident. This naturally
leads us to the question why the SRC reduces the \esym at both
sub-saturation and supra-saturation densities. The SRC affects the
\esym through several terms. First of all, because of the
momentum-squared weighting in calculating the average nucleon
kinetic energy, the isospin dependence of the HMT makes the kinetic
symmetry energy different from the FFG prediction as already pointed
out in several earlier
studies\,\cite{CXu11,CXu13,Hen15b,Vid11,Lov11,Car12,Car14}.
More specifically, within the parabolic approximation of ANM's EOS
the \esym is approximately the energy difference between PNM and
SNM. Thus, the larger HMT due to the stronger SRC dominated by the
neutron-proton isosinglet interaction increases significantly the
average energy per nucleon in SNM but has little effect on that in
PNM, leading to a reduction of the kinetic symmetry energy. 

It is worth emphasizing that we focused on effects of the SRC on the symmetry energy of uniform and cold neutron-rich nucleonic matter within the quasi-nucleon picture. It is known that at very 
low densities symmetric nuclear matter is unstable against forming clusters, such as deuterons and alphas.  For studies on the symmetry free energy of clustered matter at finite temperature we refer the readers to
refs. \cite{Joe, Typel}.

In summary, within a modified non-relativistic GHF-EDF approach and using a new momentum regulating function, we studied effects of SRC-induced HMT in the single-nucleon momentum distribution on the density dependence of nuclear symmetry energy. After re-optimizing the modified GHF-EDF by reproducing the same empirical properties of ANM, SNM and major features of nucleon optical potential at saturation density, the \esym was found to decrease at both sub-saturation and supra-saturation densities, leading to a reduced curvature $K_{\rm{sym}}$ of \esym and subsequently a smaller $K_{\tau}$ for the isospin-dependence of nuclear incompressibility in better agreement with its experimental value.  Moreover, the SRC-modified EOS and the single-nucleon potentials in ANM can be used in future transport model simulations of heavy-ion collisions to investigate SRC effects in dense neutron-rich matter in terrestrial laboratories.\\

\section{Acknowledgement}
We thank X.H. Li, W.J. Guo and X.T. He for helpful discussions in earlier studies on this topic. 
This work was supported in part by the U.S. Department of Energy, Office of Science, under Award Number DE-SC0013702, the CUSTIPEN (China-U.S. Theory Institute for Physics with Exotic Nuclei) under the US Department of Energy Grant No. DE-SC0009971 and the National Natural Science Foundation of China under Grant No. 11320101004 and 11625521, the Major State Basic Research Development
Program (973 Program) in China under Contract No. 2015CB856904, the Program for Professor of Special Appointment (Eastern Scholar) at Shanghai Institutions of Higher Learning, Key Laboratory for Particle Physics, Astrophysics and Cosmology, Ministry of Education, China, and the Science and Technology Commission of Shanghai Municipality (11DZ2260700).



\end{document}